\documentclass[twocolumn,prl,showpacs]{revtex4}
\usepackage{epsfig}

\newcommand{\la}{\langle}
\newcommand{\ra}{\rangle}

\newcommand{\Z}{{Z \!\!\! Z}}
\newcommand{\beqn}{\begin{eqnarray}}
\newcommand{\eeqn}{\end{eqnarray}}
\newcommand{\eq}[1]{(\ref{#1})}

\newcommand{\dual}{\mbox{}^{\ast}}
\newcommand{\phot}{{\mathrm{phot}}}
\newcommand{\mon}{{\mathrm{mon}}}
\newcommand{\mix}{{\mathrm{mix}}}

\begin{document}


\title{Photon propagator, monopoles and the thermal phase
   transition in 3D compact QED}

\author{M.~N.~Chernodub}
\affiliation{
ITEP, B.Cheremushkinskaja 25, Moscow, 117259, Russia}
\affiliation{Institute for Theoretical Physics, Kanazawa
University, Kanazawa 920-1192, Japan}
\author{E.-M.~Ilgenfritz}
\affiliation{Research Center for
Nuclear Physics, Osaka University, Osaka 567-0047, Japan}
\author{A.~Schiller}
\affiliation{Institut f\"ur Theoretische Physik and NTZ, Universit\"at
Leipzig, D-04109 Leipzig, Germany}


\begin{abstract}
We investigate the gauge boson propagator in three dimensional
compact Abelian gauge model in the Landau gauge at finite
temperature. The presence of the monopole plasma in the
confinement phase leads to appearance of an anomalous dimension in
the momentum dependence of the propagator. The anomalous dimension
as well as an appropriate ratio of photon wave function
renormalization constants with and without monopoles are observed
to be order parameters for the deconfinement phase transition. We
discuss the relation between our results and the confining
properties of the gluon propagator in non--Abelian gauge theories.
\end{abstract}


\date{\today}

\maketitle




Three--dimensional compact electrodynamics (cQED$_3$) shares two
outstanding features of QCD, confinement~\cite{Polyakov} and
chiral symmetry breaking~\cite{ChSB}.
With some care, it might be
helpful for the
understanding of certain non--perturbative aspects of QCD
to study them within cQED$_3$.
The non--perturbative properties of cQED$_3$ deserve interest by
themselves because this model was shown to describe some features
of Josephson junctions~\cite{Josephson} and high--$T_c$
superconductors~\cite{HighTc}.

Here, we want to elaborate
on cQED$_3$ as a toy model of {\it confinement}.
Indeed, this has been the first
non--trivial case in which confinement of electrically charged
particles was understood analytically~\cite{Polyakov}. Confinement
is caused here by a plasma of monopoles which emerge due to the
compactness of the gauge field. Other common features of the two
theories are the existence of a mass gap and of a
confinement--deconfinement phase transition at some non--zero
temperature.
According to universality arguments~\cite{universality} the
phase transition of cQED$_3$ is expected to be of Kosterlitz-Thouless
type~\cite{KT}.

In QCD$_4$, the deconfinement phase transition is widely believed
to be caused by loss of monopole {\it condensation} (for a review
see Ref.~\cite{monopole:condensation}) within the effective dual
superconductor approach~\cite{dual:superconductor}. Studying the
dynamics of the monopole current inside gluodynamics, monopole
de--condensation at the critical temperature is appearing as {\it
de--percolation}, {\it i.e.} the decay of the infrared, percolating
monopole cluster into short monopole
loops~\cite{monopole:percolation}.
This change of vacuum structure has a {\it dimensionally reduced}
analog in the $3D$ monopole--antimonopole pair binding which has been
observed in cQED$_3$~\cite{Binding,CISPaper1}.

At present, the gluon propagator in QCD$_4$ is under intensive
study. The analogies mentioned before encouraged us to study the
similarities between the gauge boson propagators in both theories.
In order to fix the role of the monopole plasma in cQED$_3$, not
just for confinement of external charges but also for the
non-perturbative modification of the gauge boson propagator, we
consider it in the confinement and the deconfined phases. On the
other hand, on the lattice at any temperature we are able to
separate the monopole contribution to the propagator by means of
eq.~\eq{eq:theta-decomp} below.

We have chosen the Landau gauge since it has been adopted in most
of the investigations of the gauge boson propagators in
QCD~\cite{CurrentQCD,Kurt} and QED~\cite{MIP,MMP}. In order to
avoid the problem of Gribov copies~\cite{Gribov}, the alternative
Laplacian gauge has been used recently~\cite{Laplacian}. The
Coulomb gauge, augmented by a suitable global gauge in each time
slice~(minimal Coulomb gauge) has been advocated both
analytically~\cite{Zwanziger} and numerically
\cite{Gribov:numerical}.

The numerical lattice results for gluodynamics show that the
propagator for all these gauges in momentum space is less singular
than $p^{-2}$ in the immediate vicinity of $p^2 = 0$. Moreover,
the results for the propagator at zero momentum are ranging from a
finite~\cite{Laplacian} (Laplacian gauge) to a strictly vanishing
~\cite{Gribov,Gribov:numerical,Zwanziger} (Coulomb gauge) value.
Recent investigations in the Landau gauge show that, beside the
suppression at $p \to 0$, the propagator is enhanced at
intermediate momenta which can be characterized by an anomalous
dimension~\cite{CurrentQCD} (see the last reference
in~\cite{CurrentQCD} for a comparison of different model
functions).

In the present letter we demonstrate that the momentum behaviour of the
photon propagator in QED$_3$ is also described
by a Debye mass and by an anomalous dimension
which both vanish at the deconfinement transition.
This mechanism can be
clearly attributed to
magnetic monopoles.
The plasma contribution is relatively easy to exhibit
by explicit calculation
and can be eliminated by {\it monopole subtraction}
on the level of the gauge fields.
The results of a study of the propagator in $SU(2)$ gluodynamics
have been interpreted~\cite{Kurt} in a similar spirit,
where $P$-vortices appearing in the maximal center gauge
were shown to be essential for the enhancement of
the Landau gauge propagator at intermediate momenta.

For our lattice study we have adopted the Wilson action,
$S[\theta] = \beta \sum_p \left( 1 - \cos \theta_p \right)$, where
$\theta_p$ is the $U(1)$ field strength tensor represented by the
plaquette curl of the compact link field $\theta_l$, and $\beta$
is the lattice coupling constant related to the lattice spacing
$a$ and the continuum coupling constant $g_3$ of the $3D$ theory,
$\beta = 1 \slash (a\, g^2_3)$. We focus here on the difference
between confined and deconfined phase. All results presented have
been obtained on lattices of size $32^2 \times 8$.
The finite temperature phase transition is known to take
place~\cite{CISPaper1,Coddington} at $\beta_c \approx 2.35$.

The Landau gauge fixing is defined by maximizing the functional
$\sum_l \cos\theta^{G}_l$ over all gauge transformations $G$. For
details of the Monte Carlo algorithm we refer to \cite{CISPaper1}.
A more complete presentation of our studies, including also a
thorough analysis of the propagator in the zero temperature case
is in preparation~\cite{in-preparation}. Details on the
implementation of Landau gauge fixing, including the elimination
of zero momentum modes and the careful control  of double Dirac
strings can be found in Ref.~\cite{MMP,in-preparation}.

We study the gauge boson propagator, $\la \theta_{\mu}(x)
\theta_{\nu} (0)\ra$, in the momentum space. The propagator is a
function of the lattice momentum, $p_{\mu} = 2 \sin (\pi k_\mu \slash
L_\mu)$, where $k_{\mu}=0,\dots,L_{\mu}\slash 2$ is an integer.
We discuss here the finite temperature case and focus on the
temporal component of the propagator,
\beqn
  D_{33}({\mathbf p}^2,0) =
  \frac{1}{L_x L_y L_z} \la \theta_3({\mathbf p},0)
  \theta_3(-{\mathbf p},0) \ra \,
  \label{eq:D_33}
\eeqn
as function of the spatial momentum, ${\mathbf p}^2 =
\sum_{\mu=1}^2 p_{\mu}^2$. We remind that at finite temperature
the confining properties of static electrically charged particles
are encoded in the temporal component of the gauge boson field,
$\theta_3$.

In order to pin down the effect of monopoles we have divided the
gauge field  $\theta_l$ into a regular (photon) and a singular
(monopole) part which can be done following Ref.~\cite{PhMon}. In the
notation of lattice forms this is written:
\beqn
  \theta = \theta^{\phot} + \theta^{\mon}\,, \quad
  \theta^{\mon} = 2 \pi \Delta^{-1} \delta p[j]\,,
  \label{eq:theta-decomp}
\eeqn
where $\Delta^{-1}$ is the inverse lattice Laplacian and the
0-form $\dual j \in \Z$ is nonvanishing on the sites of the dual
lattice occupied by the monopoles. The 1-form  $\dual p[j]$
corresponds to the Dirac strings (living on the links of the dual
lattice) which connect monopoles with anti--monopoles, $\delta
\dual p[j] = \dual j$. For a Monte Carlo configuration, we have
fixed the gauge, then located the Dirac strings, $p[j]\ne0$, and
constructed the monopole part $\theta^{\mon}$ of the gauge field
according to the last equation in \eq{eq:theta-decomp}. The photon
field is just the complement to the monopole part according to the
first equation of \eq{eq:theta-decomp}.

\begin{figure}[!htb]
  \begin{center}
    \epsfxsize=6.5cm
    \epsffile{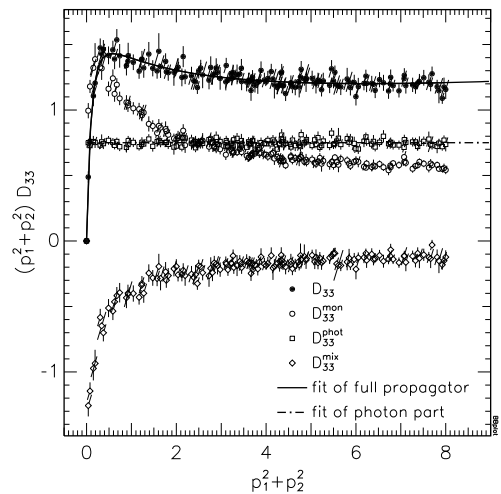}
  \end{center}
  \caption{Different contributions to the full $D_{33}$ propagator
  (multiplied by ${\mathbf p}^2$) vs
  spatial lattice momentum squared and fits as described in the
  text for $\beta=1.8$ on a $32^2 \times 8$ lattice.}
  \label{fig:propagator-parts}
\end{figure}

The photon and monopole parts of the gauge field contribute to the
propagator, $D = D^{\phot} + D^{\mon} + D^{\mix}$, where $D^{\mix}$
represents the mixed contribution from regular {\it and} singular fields.
We show the propagator for $ p = ({\mathbf p},0)$ together with the
separate contributions, multiplied by ${\mathbf p}^2$ and averaged over
the same ${\mathbf p}^2$ values, in Figure~\ref{fig:propagator-parts}
for coupling constant $\beta=1.8$.

The regular part of the propagator has perfectly the free field
form
\beqn
  D_{33}^{\phot} = \frac{1}{\beta} \frac{Z^{\phot}}{{\mathbf p}^2}\,,
\label{eq:reg}
\eeqn
at all available $\beta$.
The perturbative propagator defined
in terms of $\theta_l$ is obviously proportional to $g_3^2$, which
is taken into account by the factor $1 \slash \beta$ in eq.~\eq{eq:reg}.
The fits of the photon part of the propagator by the above expression
give the parameter $Z^{\phot}$ as a function of lattice coupling
(dash-dotted line in Figure~\ref{fig:propagator-parts} for $\beta=1.8$).

The singular contribution to the gauge boson
propagator
shows a maximum in ${\mathbf p}^2 D_{33}^\mon$
at some momentum (Figure~\ref{fig:propagator-parts}),
moving
with increasing $\beta$ nearer to $|{\mathbf p}|\, a=0$.
The mixed component
gives a negative contribution to
${\mathbf p}^2 \, D_{33}^{\mix}$, growing with decreasing momentum.
The central point of our paper is that all these contributions together
do {\it not} sum up to a simple massive Yukawa propagator.
To quantify the difference between a Yukawa--type and the actual behavior
we use the the following four--parameter model function for
$D_{33}({\mathbf p}^2,0)$,
\beqn
  D_{33}({\mathbf p}^2,0)
  = \frac{Z}{\beta} \frac{m^{2 \alpha}}
  {{\mathbf p}^{2 (1+\alpha)} + m^{2 (1+ \alpha)} } + C\,,
  \label{eq:fit3}
\eeqn
where $Z$, $\alpha$, $m$ and $C$ are the fitting parameters. This
model is similar to some of Refs.~\cite{CurrentQCD,Ma} where the
propagator in gluodynamics has been studied.

The first part of the function \eq{eq:fit3} implies
that the photon acquires a Debye mass $m$ (due to
screening~\cite{Polyakov}) together with the anomalous dimension $\alpha$.
The (squared) photon wave function renormalization constant $Z$
describes the renormalization of the
 photon wave function due to quantum
corrections. The second part of \eq{eq:fit3} represents a
$\delta$--function--like interaction in
coordinate space.

Before fitting we average the propagator over all lattice momenta at same
${\mathbf p}^2$ to improve rotational invariance.
Thus the errors entering the fits include both the variance
among the averages for individual momenta and the individual errors.
The fits were performed using standard Mathematica packages
combined with a search for the global minimum in $\chi^2 \slash d.o.f.$
To check the stability of the fits, we studied several possibilities of
averaging and thinning out the data sets, a procedure which will be discussed
elsewhere~\cite{in-preparation}.

The model function \eq{eq:fit3} works perfectly for all
${\mathbf p}^2$ and couplings $\beta$.
For $\beta \geq 2.37$ the best fit for mass parameter $m$
and anomalous dimension $\alpha$ are both consistent with zero.
Therefore we set $m=0$ and $\alpha=0$ for these values of $\beta$
to improve the quality of the fit of $Z$ and $C$.

It turns out that the inclusion of a constant term, $C$, in the
model function \eq{eq:fit3} is crucial for obtaining good fits in
the confinement phase, despite the fact that it is very small (as
function of $\beta$ the parameter $C$  decreases from $C(1.0)
=0.18(4)$ to $C(2.2)=0.009(2)$, it rapidly vanishes in the
deconfined phase). Similarly to $m$ and $\alpha$ parameters we set
$C$ to zero for $\beta \geq 2.45$, where $C$ becomes smaller than
$10^{-4}$.

\begin{figure}[!htb]
  \begin{center}
    \epsfxsize=6.5cm
    \epsffile{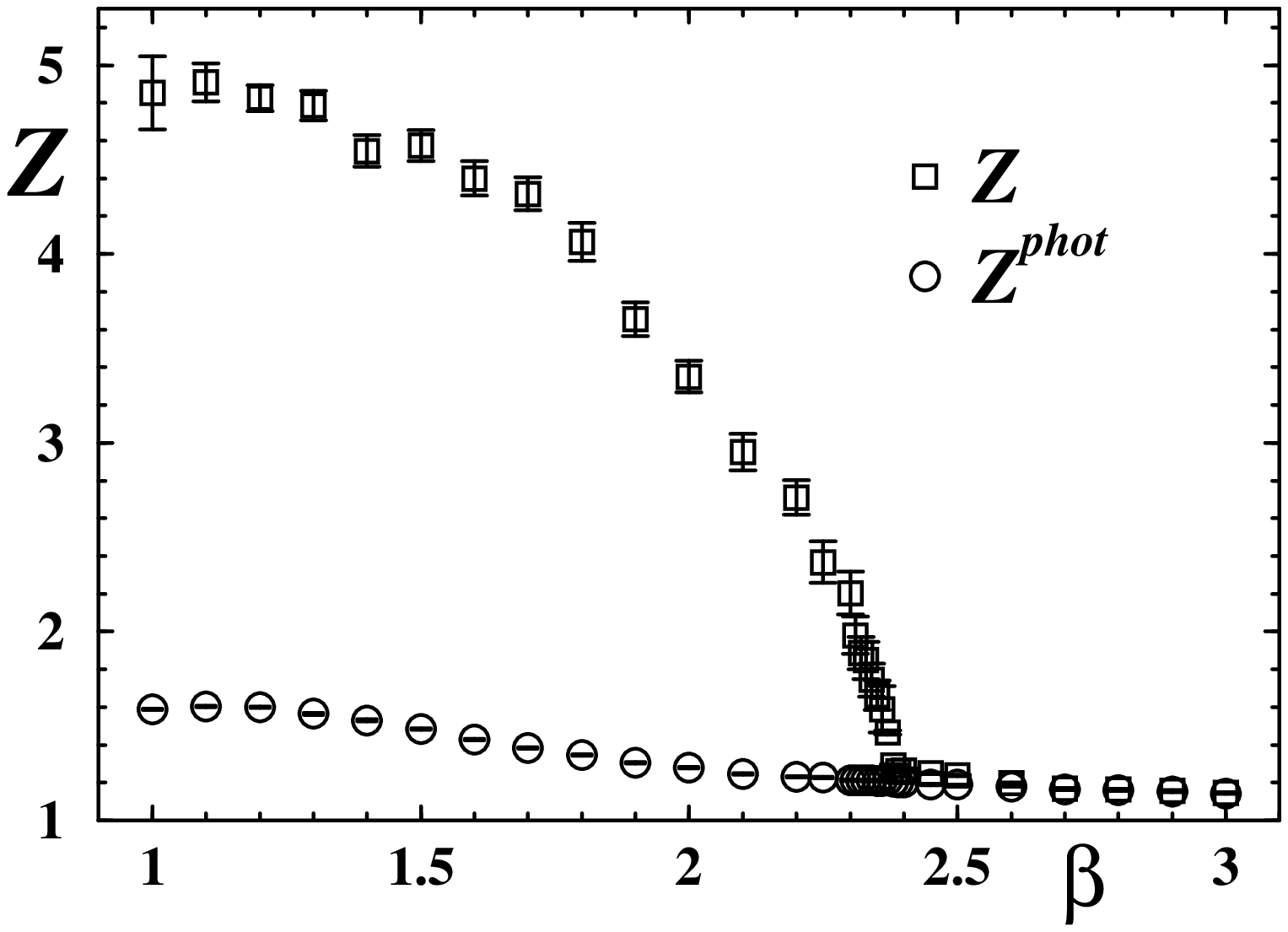}
  \end{center}
  \caption{Coefficients $Z$ of fit \eq{eq:fit3} for full propagator
   and $Z^{\phot}$ for photon contribution \eq{eq:reg} vs $\beta$.}
  \label{fig:Z-Zphot}
\end{figure}

An example of the best fit of the full propagator for $\beta=1.8$
is shown in Figure~\ref{fig:propagator-parts} by the solid line
(with $C=0.033(5)$). The parameter $Z$ distinguishes clearly
between the two phases (Figure~\ref{fig:Z-Zphot}).
It coincides with the photon part $Z^{\phot}$ (defined without
monopoles) in the deconfined phase while it is much larger in the
confined phase. This indicates that the photon wave function gets
strongly renormalized by the monopole plasma. In contrast, the
factor $Z^{\phot}$ smoothly changes crossing the deconfinement
transition at $\beta_c \approx 2.35$.

The anomalous dimension $\alpha$ also distinguishes the two
phases (Figure~\ref{fig:alpha}):
\begin{figure}[!htb]
  \begin{center}
     \epsfxsize=6.5cm
     \epsffile{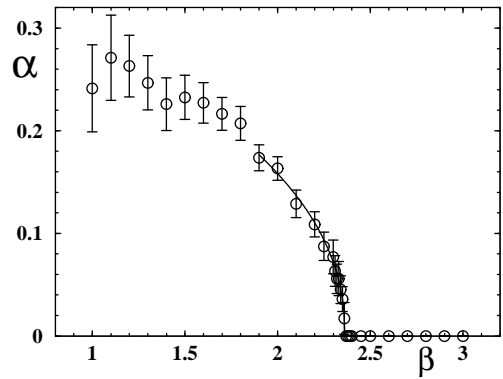}
  \end{center}
  \caption{Anomalous dimension $\alpha$ vs $\beta$ and its best
          fit near $\beta_c$ using function \eq{fitRZ}.}
\label{fig:alpha}
\end{figure}
it is equal to zero in the deconfinement phase (perturbative behaviour)
while in the confinement phase the monopole plasma causes
the anomalous dimension growing to $\alpha \approx 0.25 \ldots 0.3$.

To characterize the properties of $Z$ and $\alpha$ approaching
the phase transition we fit the excess of the ratio of $Z$'s over unity,
\beqn
  R_Z(\beta) = \frac{Z(\beta)}{Z^{\phot}(\beta)}-1\,,
  \label{eq:RZ}
\eeqn
and the anomalous dimension $\alpha$ in the following form:
\beqn
  f_i(\beta) = h_i\, (\beta^{(i)}_c -
  \beta)^{\gamma_i}\,,\quad \beta < \beta^{(i)}_c
   \,,\quad(i=\alpha,Z)\,.
  \label{fitRZ}
\eeqn
where $i=Z,\alpha$. The $\beta^{(\alpha,Z)}_c$ are
the pseudo--critical couplings which might differ on finite lattices.

\begin{figure}[!htb]
  \begin{center}
    \epsfxsize=6.5cm
    \epsffile{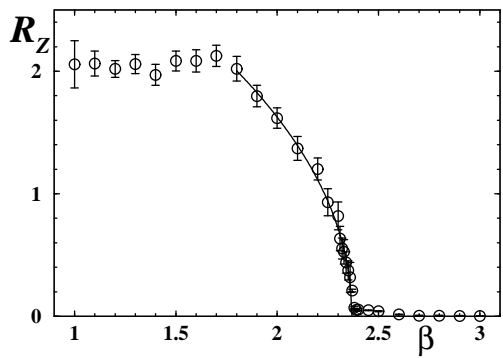}
  \end{center}
  \caption{Same as in Figure~\ref{fig:alpha} for ratio $R_Z$,
           eq.~\eq{eq:RZ}.}
  \label{fig:ratioZ}
\end{figure}

The best fits $f_{\alpha}$ and $f_Z$ are shown
in Figures~\ref{fig:alpha} and \ref{fig:ratioZ}, respectively.
The solid lines in both plots extend over the
fitting region. The corresponding parameters are presented in
Table~\ref{table:fit}.
\begin{table}[!htb]
  \begin{center}
  \begin{tabular}{|c|c|c|c|}
  i & $h_i$   & $\beta^{(i)}_c$ & $\gamma_i$ \\
  \hline
   $\alpha$ &  0.250(9) & 2.363(3) & 0.50(2) \\
        $Z$ &  2.63(7)  & 2.368(5) & 0.48(3) \\
  \end{tabular}
  \end{center}
  \caption{Best parameters for the fits \eq{fitRZ}.}
  \label{table:fit}
\end{table}
The pseudo--critical couplings $\beta^{(\alpha)}_c$ and
$\beta^{(Z)}_c$ are in agreement with previous numerical
studies~\cite{Coddington,CISPaper1} giving $\beta_c  = 2.346(2)$.
Note that the critical exponents $\gamma_i$ are close to $1 \slash
2$, both for the anomalous dimension $\alpha$ and for $R_Z$
expressing the ratio of photon field renormalization constants.

\begin{figure}[!htb]
  \begin{center}
    \epsfxsize=6.5cm
    \epsffile{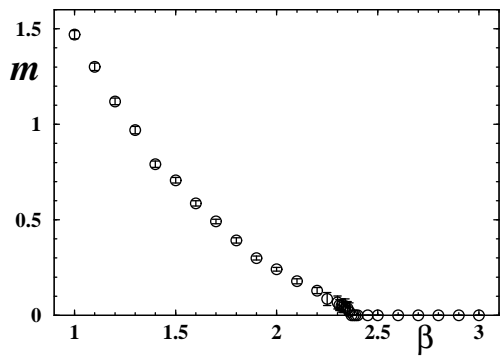}
  \end{center}
  \caption{The mass $m$ vs $\beta$.}
  \label{fig:m}
\end{figure}

Finally, the $\beta$--dependence of the mass parameter, $m$, is
presented in Figure~\ref{fig:m}. As expected, the mass scale
generated is non--vanishing in the confinement phase due to
presence of the monopole plasma~\cite{Polyakov}. It vanishes at
the deconfinement transition point when the very dilute remaining
monopoles and anti--monopoles form dipoles~\cite{CISPaper1}.

Summarizing, we have shown that the presence of the monopole
plasma leads to the appearance of a non--vanishing anomalous
dimension $\alpha>0$ in the boson propagator of cQED$_3$ in the
confinement phase.
We would hope that our observation stimulates an analytical
explanation.

At this stage of studying cQED$_3$ as a model of confinement we
conjecture that in the case of QCD the Abelian monopoles defined
within the Abelian projection may be responsible for the anomalous
dimension of the gluon propagator observed in
Refs.~\cite{CurrentQCD,Ma}. If true, a monopole subtraction
procedure analogous to that employed here would be able to
demonstrate this. We found that the anomalous dimension $\alpha$
and the ratio of the photon wave function renormalization
constants with and without monopoles, $R_Z$ \eq{eq:RZ}, represent
alternative, also non--local order parameters characterizing the
confinement phase.

\begin{acknowledgments}
The authors are grateful to P.~van Baal, K.~Langfeld,
M.~M\"uller-Preussker, H.~Reinhardt and D.~Zwanziger for useful
discussions. M.~N.~Ch. is supported by the JSPS Fellowship P01023.
E.-M.~I. gratefully appreciates the support by the Ministry of
Education, Culture and Science of Japan (Monbu-Kagaku-sho) and the
hospitality extended to him by H. Toki. He thanks for the
opportunity to work in the group of H. Reinhardt at T\"ubingen and
for a CERN visitorship in an early stage of this work.
\end{acknowledgments}

\end{document}